\documentclass{epl}
\usepackage{graphicx}
\usepackage{latexsym}
\usepackage{amsmath}
\title{Interaction effects between impurities in low dimensional spin-1/2
antiferromagnets}
\shorttitle{Interaction effects between impurities}
\author{Fabrizio Anfuso\inst{1} \and Sebastian Eggert\inst{1,2}}
\institute{
\inst{1} Dept. of Physics,
Chalmers Univ. of Technology, 
S-412 96 G\"oteborg, Sweden\\
\inst{2} Dept.~of Physics, Univ.~of Kaiserslautern, 
D-67663 Kaiserslautern, Germany}
 \pacs{75.10.Jm}{Quantized Spin Models}
 \pacs{75.30.Hx}{Magnetic impurity interactions}
\begin{document}

\maketitle

\begin{abstract}
We are considering the interplay between several non-magnetic impurities in the spin-1/2
Heisenberg antiferromagnet in chains, ladders and planes
by introducing static vacancies in numerical quantum Monte Carlo simulations.
The effective potential between two and more impurities 
is accurately determined, which gives a direct measure of the quantum correlations in the systems.
Large effective interaction potentials are an indication of strong quantum correlations in the system
and reflect the detailed nature of the 
valence bond ground states.  In two-dimensions (2D) the interactions are smaller, but can still be
analyzed in terms of valence bonds.
\end{abstract}
\section{Introduction} \label{intro}
The interplay between impurities in antiferromagnetic backgrounds has become an important topic ever
since the discovery of high-temperature superconductivity.   The interaction energy between mobile 
holes has been examined with a variety of numerical and analytical methods for t-J and 
Hubbard models
in order to get a better insight into the pairing mechanism in
the HiT$_c$ compounds\cite{Trugman,Poilblanc,Bulut2,White}.  Equally interesting is the 
effect 
of static impurities\cite{Bulut,Sushkov,Song,Nagaosa,behre} which for example are known to 
induce spin-1/2 degrees in ladders, that interact to form gapless excitations in otherwise
gapped systems\cite{Mikeska,Sigrist}.  For static vacancies one of the first studies was 
performed by Bulut et al.~with linear spin-wave approximation
and numerics on small lattices\cite{Bulut}, establishing a nearest neighbor 
attraction and an interesting enhancement of the local quantum correlations. 

We are now considering several static holes (vacancies) 
in low dimensional antiferromagnets like chains, ladders and planes on larger lattices, 
in order to calculate their impurity and interaction energies in the
limit of zero temperature.
The energies are changed because nearest neighbor correlations 
$\langle\mathbf{S}_{i}\cdot \mathbf{S} _{j}\rangle$ are altered
around the vacancies which in turn visualize the quantum correlations on the lattice.
Therefore the interaction energy between two or more impurities gives direct information on 
the quantum correlations by showing the underlying local 
valence bond order as will be explained for a variety
of configurations in the antiferromagnetic spin-1/2 Heisenberg model
$
H=J\sum_{\langle i,j \rangle}\mathbf{S}_{i}\cdot \mathbf{S}_{j} $
where $\langle i,j \rangle$ denotes nearest neighbor sites on chain, ladder and square 
lattices.  

In order to determine impurity contributions to the total energy and the effective 
potentials between vacancies we have to sum up all bond strengths 
in the system and subtract the reference energy of a pure system.
Since it is necessary to subtract two 
large extensive quantities, the energies $E$ for each configuration must be determined to at least six 
significant digits in our case.  For practical purposes this limits the system sizes for 
determining the impurity
energies to a periodic $16\times 16$ lattice for the 2D geometry. 
The quantum Monte Carlo program we developed uses the loop algorithm in a 
single cluster variety implemented in continuous
time\cite{Evertz,Evertz1,Wolff,Beard}, which gives efficient and fast updates 
even at very low temperatures.
From now on all temperatures and energies are 
given in units of $J=1$.

\begin{figure}
\begin{center}
\includegraphics[width=70mm]{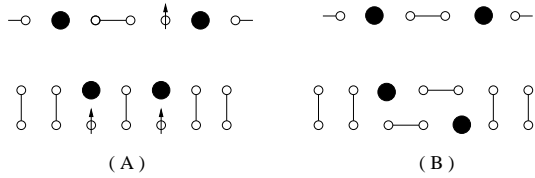}
\end{center}
 \caption{Different configurations of valence bonds around impurities.  In cases A it is not possible to construct 
 a simple valence bond state around the impurities, so the energy cost is higher.  In the cases B the energy is lowered
 because quantum correlations on sites 
 between the impurities are enhanced by the valence bonds as drawn.}
 \label{VBS}
 \end{figure}

In order to understand the meaning of the interaction energy between impurities, let us first
consider the extreme limit of the absence of all quantum fluctuations like in 
the Ising model, where the ground state is characterized by a purely classical N\'eel 
order.  
Substitutional impurities remove antiferromagnetic bonds, but do not change the ordered ground state and 
cannot feel an effective potential between each other 
except for a nearest neighbor attraction, since 
more antiferromagnetic bonds are present when impurities are placed on neighboring sites.
However, in systems with an entangled quantum ground state 
there is a possibility that impurities can have an effective potential over 
several lattice sites, since the quantum correlations are strongly altered by the impurities (while classical order is not).  
Such quantum correlated systems can often be 
best described in terms of a resonating
valence bond basis or a valence bond solid.  
In that case, there is a higher  energy cost if impurities are placed in such a way 
that the valence bonds (singlets) cannot be accommodated between the vacancy sites, 
while the effective potential may be attractive
if resonating valence bonds fit exactly between the impurities.  
In effect the vacancies prune the possible singlet 
formation analogous to the ideas of Martins et al.\cite{Martins,Martins1} 
where the antiferromagnetic order around single impurities was discussed.
Figure \ref{VBS} shows typical configurations
for attractive and repulsive configurations in chains and ladders, for which
resonating valence bonds form a good description of the systems.
The effective potential between vacancies is therefore 
a direct indication both of the configuration and the amount of 
quantum correlations in the system as compared to 
a classical N\'eel state.

\section{Vacancies in spin-chains}
\begin{figure}
\begin{center}\includegraphics[width=85mm]{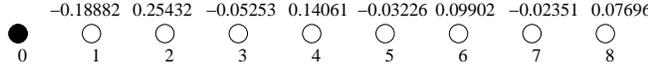}\end{center}
\caption{The exact impurity-impurity interactions in the 1D lattice.}
\label{1dfig}
\end{figure}

In the one-dimensional case, vacancies simply cut the infinite chain, so that the impurity 
interaction potential can easily be determined from the ground state 
energy of the finite segment between the impurities.
Finite Heisenberg spin chains are extremely well studied so that 
the interaction potential can be calculated analytically 
up to higher order corrections.  
It is useful to approximate the ground state energy $E_{\rm gs}(L)$ 
for a finite chain of length $L$ in the form 
\begin{equation}
E_{\rm gs}(L)=\epsilon_0L+ E_s +\frac{b}{L} + {\cal O}\left(\frac{1}{L \ln(L)}\right),
\label{conformale}
\end{equation}
where $\epsilon_0 = 1/4-\ln 2$ is the thermodynamic limit of the energy per site and 
$E_s  = \frac{\pi-1}{4} -\frac{\ln 2}{2} \approx 0.1888$ is the surface energy 
for open boundary conditions\cite{batchelor}. The single impurity energy is 
given by the difference of total energy of the pure system $E_0$ and of a system with 
one impurity $E_1$
\begin{equation}
\Delta E = E_1 -E_0  = E_s-\epsilon_0 \approx 0.632.
\label{delta}
\end{equation}
This impurity energy $\Delta E$ is smaller than the energy cost of two 
missing links $-2\epsilon_0 \approx 0.8863$, because the quantum correlation are changed in the 
ground state (which would not be the case in an Ising-like system).
The effective impurity-impurity potential $U(x)$ can be extracted by
comparing the total energy $E{(x)}$ in a given configuration to a 
reference-energy $E{(\infty)}$, corresponding to two vacancies very far apart.   
Equivalently, we can also define the reference energy as the extra energy 
cost of two single impurities $E{(\infty)} = E_0 + 2(E_1-E_0)$.
Accordingly the surface energy and energy per site cancel, 
and the leading contribution comes from 
from the $1/L$ term in eq.~(\ref{conformale}), which is known from conformal field theory 
to be $b= -\pi^2/48$ for open chains 
with even number of sites\cite{Blote}.  However, for an odd number of sites 
the ground state quantum numbers correspond 
to an excited state in terms of the field theory\cite{karbach} and for odd  
open chains we have \cite{suzuki}
$b=5 \pi^2/48$.  We therefore find 
\begin{equation}
U(x)=E{(x)}-E{(\infty)} \approx (-1)^x \frac{\pi^2}{16 x} + \frac{\pi^2}{24 x}, 
\label{pot}
\end{equation}
which is an alternating potential with a relatively large magnitude and
a slow powerlaw decay in agreement with the intuitive picture in fig.~\ref{VBS}. 
The results for shorter distances can easily be calculated exactly and are 
shown in fig.~\ref{1dfig}, which only slowly approach eq.~(\ref{pot}) because
of the logarithmic corrections in eq.~(\ref{conformale}).  
Interestingly, the nearest neighbor attraction is not very large, 
which comes from the fact that not very
many quantum valence bonds are enhanced by this configuration compared to two isolated impurities.  
On the other hand,  the repulsion for $U(2)$ is even stronger in magnitude 
since this placement of vacancies disrupts the valence bond 
configuration very strongly.  This is generally true for an odd number of 
sites between the vacancies as is intuitively clear from fig.~\ref{1dfig}.

\begin{figure}
\begin{center}\includegraphics[width=75mm]{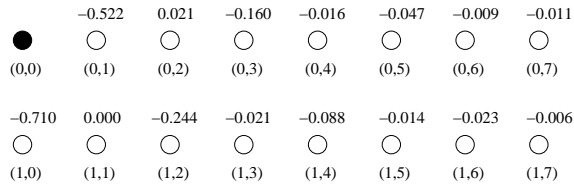}\end{center}
\caption{Potential $U(x,y)$ between static holes for a two-leg ladder of size $2\times40$ with $\beta=15$ calculated 
as $E_{(x,y)}-E_{2,20}$.}
\label{ladder}
\end{figure}

\section{Vacancies in spin-ladders}
An equally interesting example of a quantum dominated ground state is the two-leg ladder which 
can locally be described by a valence bond solid but with short-range correlations due to an energy gap.  
 Placing a single impurity into this system is already rather disruptive to the ground state, 
which is reflected by the relatively high energy cost of  $\Delta E \approx 1.215$.  However,
this energy is almost completely accounted for by the missing energy bonds along the legs
 $\epsilon_{\rm leg} \approx 0.350$ and across 
the rungs $\epsilon_{\rm rung} \approx 0.455$, as if the ground state configuration of
the entire system is not changed except for the local extraction of the missing bonds.
The numerical results for the effective impurity-impurity potential of  a $2\times 40$
ladder system are presented in fig.~\ref{ladder}.

The strongest potential is found for the nearest neighbor attraction, which is stronger 
across the rungs than along the legs, in accordance with the different energies per
bond $\epsilon_{\rm leg} < \epsilon_{\rm rung}$. 
For larger distances a staggered potential can be observed, which 
is however predominantly negative.  This means that in most cases 
the valence bond state can be partially repaired between two close vacancies,
even for two impurities on the same sublattice.
Interestingly, the zig-zag pattern of attractive energies 
along the sites on the opposite sublattice
$(1,0), (0,1),(1,2),(0,3)....$ fits well to an exponential decay 
$U(r) \sim - 1.05 \exp(-0.632 r)$  with geometrical distance $r$.  This means that the
potential decay length $\xi_U\sim 1.58$ is almost exactly half 
of the decay length for correlations $\xi \sim 3.19$\cite{noack} or for effective
spin interactions $\xi \sim 3.1$\cite{Mikeska}.  The overall magnitude of the 
potential is relatively large as expected for a system with strong quantum correlations, but 
it remains short range.

 \begin{figure}
\begin{center}\includegraphics[width=80mm]{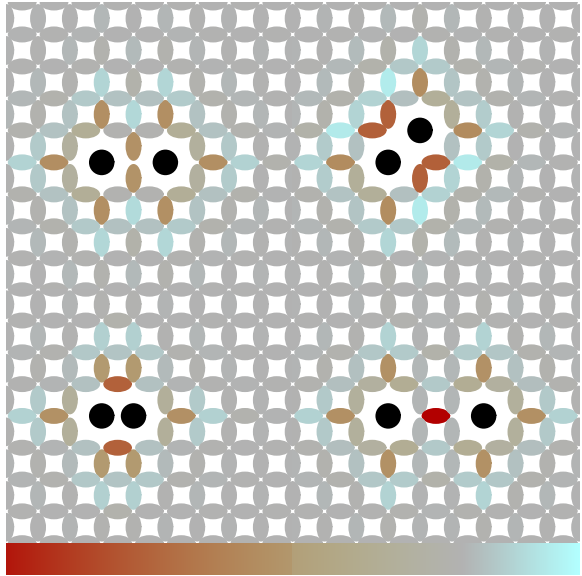}\end{center}
 \caption{(Color online) Different configurations of impurities in the 2D 
 Heisenberg antiferromagnet and the corresponding 
 bond strengths on a scale ranging from $-0.4365$ to $-0.312$.  In the upper two cases  
 the valence bonds are disrupted in a relatively large area around the vacancies,
 so that the energy cost is higher.  
 In the last case the energy is lowered
 because quantum correlations on sites between the impurities are enhanced.}
 \label{bonding}
 \end{figure}
\section{Vacancies in the 2D Heisenberg antiferromagnet}
In contrast to chains and ladders the 2D Heisenberg antiferromagnet is not dominated by 
strong quantum correlations.  Instead it is believed to 
be well described by spin waves on top of a N\'eel ordered state.   However, some 
valence bond character may still remain, especially on shorter distances, which will be
revealed by the study of the impurity-impurity interactions.   As argued above, a classical 
antiferromagnetic order does not imply any interaction energies, so that the 
impurity-impurity potential will give information about the magnitude and nature of the 
remaining valence bond character in the 2D antiferromagnet.

Indeed we find a pronounced change of the correlation strengths 
$\langle \mathbf{S}_{i}\cdot \mathbf{S}_{j}\rangle$ of the nearest neighbor bonds as is shown
for some typical arrangements in
fig.~\ref{bonding} at low temperatures from our numerical simulations. 
A minimum value of $-0.75$
would correspond to a singlet and $\epsilon_0 \approx -0.3347$ is the bulk value.
The valence bond character closest to the impurities
is always enhanced\cite{Martins,Martins1}, but this can lead to a weakening of 
other bonds. The sum of all bond strengths gives the total energy and there may 
be a net energy cost (or gain), depending on the configuration. 
For example, if a vacancy is placed at $(0,0)$ and a second vacancy at $(1,1)$ or $(0,2)$ the
valence bond order is disturbed locally so that we see a net energy cost.  For a second
vacancy at $(0,3)$ on the other hand the valence bond between the impurities is so strongly 
enhanced as shown in fig.~\ref{bonding} that this configuration is energetically 
relatively favorable.
All simulations of fig.~\ref{bonding} were done on a $16\times 16$ lattice at $\beta J = 30$, 
but are shown in one single plot for convenience.

\subsection{One vacancy}
\begin{figure}
\begin{center}
\includegraphics[width=65mm]{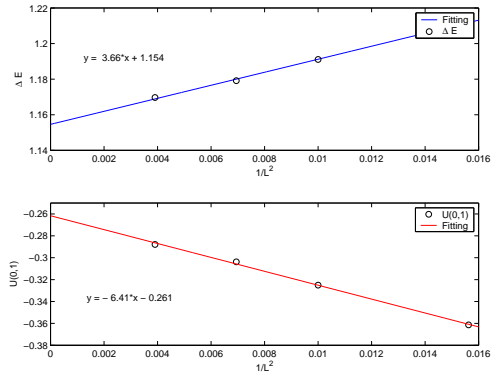}
\end{center}
\caption{(Color online) The finite size scaling of $\Delta E$ and  $U(0,1)$ in an $L \times L$ lattice.  The impurity quantities $\Delta E$ and $U(0,1)$
converge with $1/L^2$ and are derived 
as a difference of extensive quantities and therefore have larger errorbars.  }
\label{scaling}
\end{figure}

For the single impurity case, we find an energy cost of 
$\Delta E = E_1 -E_0 \approx 1.154$ which is 
smaller than the energy cost of removing four bonds 
$\Delta E +4\epsilon_0 \approx -0.18$, where $\epsilon_0 \approx -0.3347$.
Those results compare well with linear spin-wave estimates of Bulut et al.\cite{Bulut} 
where it was observed that large lattices are necessary 
for convergence, which is an indication that the
system is affected by the impurity over relatively large distances.  In fig.~\ref{scaling} 
the scaling behavior for 
the most important quantities are shown, indicating that $\Delta E$ has converged to almost
99\% for a $16 \times 16$ lattice.   
Typically, impurity quantities converge to the thermodynamic limit with $1/L^2$ 
while $\epsilon_0$ converges with $1/L^3$ on an $L\times L$ lattice\cite{Bulut}.

\subsection{Two vacancies}
\label{twopot}
In order to determine the impurity-impurity potential 
we place the first vacancy at $(0,0)$ and a second one in several different positions $(x,y)$.  The  
total energy of the system $E_{(x,y)}$ is then calculated numerically in each case.
According to eq.~\ref{pot} we define the effective impurity-impurity potential 
$U(x,y) =  E_{(x,y)}-E_{(\infty,\infty)}$.
 For the reference energy it turns out that a vacancy at maximum distance is already a perfect
approximation within numerical accuracy, i.e. $E_{(\infty,\infty)} \approx E_{(8,8)}$, which can also 
be determined with a lower statistical error than the extra energy cost of two single impurities
$E_{(\infty,\infty)} = E_0 + 2(E_1-E_0)$.

\begin{figure}
\begin{center}
\includegraphics[width=60mm]{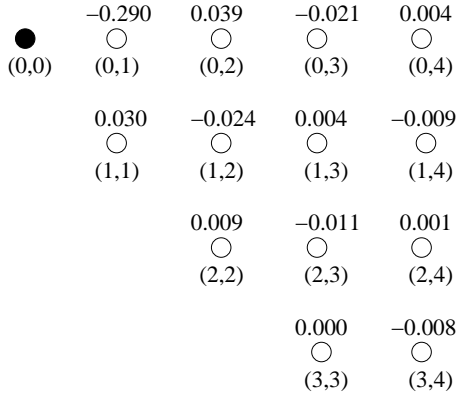}
\end{center}
\caption{Potential $U(x,y)$ between static holes in a 2D antiferromagnet on a $16\times 16$ lattice. }
\label{energy2d}
\end{figure}

Figure \ref{energy2d} shows the results of our numerical analysis for $U(x,y)$ within
our statistical error of $\pm 0.002$ on a $16\times 16$ lattice at $T=J/30$.
Not surprisingly the dominant contribution is given by an attractive nearest neighbor interaction, 
which is due to the fact that one more antiferromagnetic bond is present for this configuration. 
This is comparable to the Ising model, but here   
the value for $U(0,1)$ is only about 80\% of the average energy per bond $\epsilon_0$, indicating that 
quantum correlations account for the difference of 20\%.
The numerically extrapolated binding energy of $U(0,1) = -0.261$ for an infinite system 
is in good agreement with the linear spin-wave theory (LSW) result\cite{Bulut}
$
U_{\textrm{LSW}}(0,1)=-0.2666.
$

The quantum correlations become even more evident for larger distances which show a staggered
effective potential. As explained above, this staggered 
potential is not caused by a N\'eel antiferromagnetic order, but is an indication of how 
efficiently valence bonds can be arranged around the impurities as already shown in fig.~\ref{bonding}.
In particular, there is a higher energy cost when the second impurity is placed at $(0,2)$ or $(1,1)$, which are 
disruptive configurations for regular valence bond arrangements.
If an impurity is placed at $(0,3)$ on the other hand, a valance bond can be formed very 
efficiently between the sites
$(0,1)$ and $(0,2)$ so that in this case the enhanced quantum correlations lead to a gain in total energy and an 
effective attractive potential.  Continuing in this way an effective staggered potential is formed. 
Although the potential is small, it is not necessarily short range.  
The relatively slow scaling of $U(0,1)$ with system size and the fact that 
the bond strengths in 
fig.~\ref{bonding} are altered over a large area 
indicate that vacancies can affect the system over a long range.
However, since the effective potential for distances larger than one is always 
much smaller than typical kinetic energies
of mobile holes would be, it is unlikely that directly magnetically mediated impurity-impurity 
potentials could significantly contribute to the pairing mechanism in high temperature 
superconductors except for nearest neighbor interactions.

\subsection{Three vacancies}
\label{manypot}
If a third vacancy is now placed in the system, it will interact with the other two 
vacancies with the same
mechanism as before, namely as a probe of which possible valence bonds have been pruned.  
This means that the three body potential can in general not be 
written as a sum of two body 
potentials.  For comparison we introduce a three body correction $U_{\rm corr}$ for impurities
placed at $(0,0)$, $(x_1,y_1)$ and $(x_2,y_2)$
\begin{equation}
 U(x_1,y_1;x_2,y_2)  = E_{x_1,y_1;x_2,y_2} - E_\infty 
 = U(x_1,y_1) + U(x_2,y_2) + U(x_1-x_2,y_1-y_2) 
 + U_{\rm corr}, 
 \label{threepot}
\end{equation}
where $E_\infty$ is the 
configuration energy of three impurities infinitely apart from each other.  In our case we can 
approximate $E_\infty \approx E_{0,8;8,8}$.
The correction term $U_{\rm corr}$ is a measure
of how disruptive the total three impurity configuration is. If for example the third 
impurity is placed on a valence
bond that is strongly enhanced by the two first impurities combined, we expect an additional penalty.
In table \ref{threevacancies} the results for the most interesting configurations are shown.
The dominant contribution comes again from the nearest neighbor attraction, which  
is just a trivial count of
how many antiferromagnetic bonds are present.  However, the correction $U_{\rm corr}$
is found to be of the same order as the longer range potentials and other quantum corrections.  
Therefore, many body 
effects are essential when describing the interaction.  In particular, it is more efficient 
to place three vacancies 
in a line $(0,0),(0,1),(0,2)$ than placing them at an angle $(0,0),(0,1),(1,0)$, 
because valence bonds can form 
more efficiently along straight edges (note, that we would get the opposite 
statement if there was no three body 
correction $U_{\rm corr}$).

\begin{table}
\begin{center}
\begin{tabular}{|c|c|c|}
\hline
 $(x_1,y_1),(x_2,y_2)$ & $U(x_1,y_1;x_2,y_2)$ & $U_{\rm corr}$\\
\hline
$(0,1),(0,2)$ & -0.573 & -0.032\\
\hline
$(0,1),(1,0)$   & -0.531  & 0.019\\
\hline
$(0,1),(0,3)$ & -0.259  & 0.013 \\
\hline
$(0,1),(1,2)$   & -0.287  & -0.003\\
\hline
$(0,1),(2,1)$  & -0.268  & 0.007 \\
\hline
$(1,1),(2,0)$ & 0.105  & 0.006\\
\hline
$(1,1),(2,2)$ & 0.066  & -0.003\\
\hline
$(2,0),(2,2)$ & 0.084  & -0.003\\
\hline
$(0,8),(8,8)$  & 0 & 0 \\
\hline
\end{tabular}
\caption{The potential for three vacancies at  $(0,0)$, $(x_1,y_1)$ and $(x_2,y_2)$
compared to the sum of  two body potentials
$U_{\rm corr} = U(x_1,y_1;x_2,y_2) -U(x_1,y_1) - U(x_2,y_2) - U(x_1-x_2,y_1-y_2) $ in a 16x16 lattice for  $\beta J= 30$.
The last digit is uncertain.}
\label{threevacancies} 
\end{center}
\end{table}

\section{Conclusions} \label{concl}
In conclusion we have analyzed the interplay of several vacancies in antiferromagnetic
spin-chains, spin-ladders, and planes in the low temperature limit.
For chain and ladder systems large effective potentials can be found, which is 
a direct indication of their well-established strongly entangled ground states over long and short
ranges, respectively.
In the 2D case, a  small effective potential between two static vacancies 
can be explained in terms of a small  
quantum valence bond character on top of the N\'eel order.  
The corresponding energies are of the 
order of 10-20\% of one bond strength, which gives 
an indication of the size of the quantum corrections in this
system.  Despite the smallness of the potential,  the bonds are affected over
a long range around the impurity.  
The potential is not a simple two body effect, which is in agreement with 
the assumption that it originates from possible valence bond arrangements (and the pruning of them).

\acknowledgments
We are very thankful for helpful discussions with Mats Granath, Holger Frahm, Henrik 
Johannesson, and Matthias Vojta.  This project was in part supported by the 
Swedish Research Council.  Computer time was allocated through Swegrid grant SNIC 011/04-15.

\end{document}